# Built-in potential and band alignment of matter

Duk-Hyun Choe[1], Damien West[1], Shengbai Zhang[1,2*]

[1]Department of Physics, Applied Physics & Astronomy, Rensselaer Polytechnic Institute, Troy, NY 12180, USA.

[2]Beijing Computational Science Research Center, 10 E. Xibeiwang Road, Beijing 100193, China.

*Correspondence to: zhangs9@rpi.edu.

**The built-in potential is the interfacial potential difference due to electric dipole at the interface of two dissimilar materials. It is of central importance to the understanding of many phenomena in electrochemistry, electrical engineering, and materials science because it determines the band alignment at the interfaces. Despite its importance, its exact sign and magnitude have generally been recognized as an ill-defined quantity for more than half a century. Here, we provide a universal definition of the built-in potential. Furthermore, the built-in potential is explicitly determined by the bulk (i.e., innate) properties of the constituent materials when the system is in electronic equilibrium, while the interface plays a role only in the absence of equilibrium. Our quantitative theory enables a unified description of a variety of important properties in all types of interfaces, ranging from work functions and Schottky barriers in electronic devices to open circuit voltages and electrode potentials in electrochemical cells.**

Understanding matter by its innate properties is one of the most fundamental challenges of science. In early-to-mid 20th century, physicists and chemists have thus proposed numerous innate properties such as work function (*1,2*), electron affinity (*3*), ionization potential (*3*), redox potential (*4*), and deformation potential (*5*), to name a few. The hope is to use such simple pictures to understand the complex nature of multi-component systems. This philosophy, however, seems to have serious limitations (*6*). In particular, when two dissimilar materials are brought into contact at the interface, countless studies indicate that non-innate properties, specific to particular multi-component system, have overwhelming effects on the experimental outcome. While this suggests a failure of the constructionist approach, to reach such a conclusion requires solid knowledge of the innate properties of the constituent systems which are often ill-defined or at least ambiguous.

Many intriguing phenomena in nature occur at interfaces. Understanding their behavior is of great interest not only for the study of exciting physics such as highly conducting electrons at the interfaces between insulators (*7-10*), but also for their direct relevance to virtually all device applications – as coined by Herbert Kroemer (*11*), "the interface is the device." A basic question in these studies is the nature of the charge transfer or variation at the interfaces that causes the built-in potential between two dissimilar materials. The built-in potential exists in any type of interfaces (*12-14*), and it determines the energy band alignment between the materials which relates to a number of fundamental properties in surface/interface science (*14-18*), such as work functions, electrode potentials, open circuit voltages, Schottky barriers, and band offsets. Despite more than 50 years of investigation, however, it has not been possible to determine the *exact* sign and magnitude of the built-in potential (*13-17*). The absence of a clear understanding of the built-in potential has led to the introduction of numerous terminologies and definitions depending on the field of sciences and/or purposes, such as Galvani potential difference, interfacial potential difference, and interface dipole. A rigorous definition of the built-in potential is a prerequisite for a quantitative description of many interfacial phenomena and promises a fundamental breakthrough in our understanding of electronic/electrochemical devices (*15,17*).

A large amount of effort during the last several decades for elucidating the built-in potential can be categorized into two schools of thought. In the first school, *the built-in potential is described by the properties of the individual constituent materials*. The key question, however, is what specific property should we consider? In electrochemistry, the built-in potential (or the interfacial potential difference) is often described by a rather abstract concept called "Galvani potential" without a rigorous theoretical definition (*13*). In device physics, the work function has often been considered (*12*) but such a property is sensitive to surface conditions and fails to explain the open circuit voltages in electrochemical circuits (*13*). In fact, such constructionist approaches have led to many notorious failures in interface science, including Schottky barriers and band offsets (*15-18*). In the second school, *the built-in potential is specific to a particular interface*. In this regard, charge density is commonly used to investigate the built-in potential,

$$\Delta \psi = \frac{e}{\varepsilon_0} \int z \Delta \overline{\rho}(z) \mathrm{d}z, \qquad (1)$$

where $\Delta \overline{\rho}(z)$ is the difference in planar averaged charge densities before and after charge relaxation, and the quantities $e$ and $\varepsilon_0$ have the usual meanings. In device physics, $\Delta \overline{\rho}(z)$ is often described as a step-function and, accordingly, the built-in potential is subsequently measured from the capacitance-voltage measurement (*12*). However, charge density-based methods have a fundamental limitation because there is no clear initial basis of comparison for determining $\Delta\bar{\rho}(z)$. The definition of $\Delta\bar{\rho}(z)$ at the interface has therefore remained a contentious issue (*15-17,19-21*). Alternately, one could seemingly define the built-in potential based on the electrostatic potential, since it is simply a change in the offset of the average electrostatic potentials in dissimilar materials, but herein lies a problem. The average electrostatic potential of a bulk solid is *itself* ambiguous (*22,23*). For instance, explicit integration of the charge density to determine the average electrostatic potential, $V_0$, is a conditionally convergent quantity whose value depends explicitly on how the integral is taken. The notion of a "bulk average electrostatic potential" is therefore typically abandoned. Since both schools have serious limitations, the built-in potential remains to be an ambiguous property despite its long history of being used to describe the band alignment at interfaces (*12-18*).

In this work, we present a universal definition of the built-in potential that allows a unified quantitative description of the properties at *any* type of heterojunction interfaces. Based on the theory, we establish a new school of thought on the built-in potential: *the built-in potential is determined by the bulk properties of the constituent materials, but only if the system is allowed to reach electronic equilibrium.* As pictorially shown in Fig. 1, the key to our finding is identifying a common energy reference among dissimilar *bulk* materials (Fig. 1C). The magnitude of the built-in potential is then explicitly given by the electrostatic potential (Fig. 1D), rather than, as generally thought, the charge density (Eq. 1). Using metal/metal, metal/semiconductor, and metal/aqueous interfaces as examples, we show our general theory of built-in potential provides new perspectives into the study of interface science.

**Unrelaxed solid surface and solid/vacuum interface**

We begin by identifying innate properties of solids. We consider an unrelaxed surface (i.e., ideal surface) which is defined as a solid surface without surface charge or ionic relaxation. Determination of the ideal surface enables a decomposition of the total potential barrier at the solid surface, *D*, into the innate (i.e., bulk) and surface contributions (Figs. 2A,B): $D = B + S$, where $B$ is the innate surface dipole associated with the bulk quadrupole and $S$ is the surface dipole produced by the charge relaxation. Considerable efforts have been made to model the ideal surfaces based on charge densities, such as using superposition of atomic charges (*19,24,25*), charge truncations at the bisected plane (*20*), and Wigner-Seitz cells (*21*). It is now widely believed that the ideal surface cannot be uniquely defined since it depends on how the boundary between the solid and vacuum is chosen before the relaxation (*15-17*). Despite these concerns, we demonstrate that there exist a unique way of defining $B$ and hence the ideal surface for a given solid surface. To avoid the fundamental limitations of the charge density-based methods, here we focus on the electrostatic potential. While the absolute average potential (independent of consideration of any surface) of bulk solid is ill-defined (*23*) – when a surface is specified, one can unambiguously define the average potential of the crystal. With this information, we can insert vacuum into the bulk in a way that the surfaces have no effect on the electrostatic potential of the non-vacuum regions of the bulk (see fig. S1). We term this inserted vacuum the "*ideal vacuum*". The above process is pictorially shown in Fig. 1. Since the ideal surface does not involve any electronic/ionic relaxation, the potential relative to the ideal vacuum reflects the bulk property instead of the solid/vacuum interface typical of work function calculations. One of our key findings is the relationship $V^{(m)} = V_{iv}$ (proof is given in Supplementary Materials), where $V^{(m)}$ is a maximum value of the planar averaged electrostatic potential and $V_{iv}$ is the potential of the ideal vacuum or ideal vacuum level (see Fig. 2A). For an ideal surface, $B$ is then defined entirely by the bulk properties, $B = V^{(m)} - V_0$.

Based upon the above decomposition, we examine the work function of a solid surface which can be generally described by two terms: $\varphi = -\mu + D$ (25,26), where $\mu$ is the chemical potential of electrons (Fig. 2B). The work function of an ideal surface is then (Fig. 2A)

$$\psi = -\mu + B, \qquad (2)$$

which represents the work required to remove an electron from the solid to vacuum before the surface relaxations take place. Throughout this paper we shall name this quantity "innate built-in potential", which only depends on the properties of bulk. In analogy, the concept of ideal surface can readily be extended to innate electron affinity and ionization potential of semiconductors by replacing $\mu$ in Eq. 2 with the band edge positions relative to $V_0$. It is important to realize that $\psi$ in general depends on the surface termination (fig. S2) and orientation (fig. S3) of the surface, which is a direct consequence of the multivalued nature of the bulk quadrupole of a solid. The lack of such understanding of the innate properties of materials is the main reason the constructionist view has languished for so long in interface science. The multivalued nature of $B$ is indeed necessary in order to describe possible interactions at the interfaces between two materials. In consideration of a particular interface, however, the innate properties associated with the interface are no longer ill-defined when the orientations and surface terminations of the two materials are given.

**Solid/solid interface**

Here, we imagine the interface formation in two steps: (i) the creation of an ideal A/B interface, which consist of two ideal surfaces of material A and B (Figs. 1A-C); (ii) allowing the electrons and ions to relax to find the ground state (Fig. 1D). Comparison of the average electrostatic potential of the systems in the above two steps yields a potential shift arising from the charge transfer, which is the very definition of the built-in potential. Surprisingly, we find that the built-in potential is simply defined as the difference between $V^{(m)}$ in bulk regions of the two materials,

$$\Delta \psi_{A|B} = V_B^{(m)} - V_A^{(m)}. \qquad (3)$$

This is because $V_A^{(m)}$ and $V_B^{(m)}$ for the two ideal surfaces are aligned to the same energy, namely the *ideal vacuum*, before the charge transfer (Figs. 1,2A). Thus, the built-in potential is embedded in the electrostatic potential *per se*, irrespective of the ambiguity in the boundary between the two materials. Equation 3 can be applied to *any* type of interface, including grain boundary, metal/semiconductor, metal/aqueous, aqueous/aqueous, and even polar/non-polar interfaces (see the Supplementary Materials).

Corollary of Eq. 3: in *any* type of heterojunctions in electronic equilibrium, the built-in potential, $\Delta \psi_{A|B}$, is determined entirely from the bulk properties of the material A and B. In this case the interface charge transfer takes place in order to align the Fermi level. One can then derive the following expression using the Fermi level alignment and Eq. 3:

$$\Delta \psi_{A|B} = \psi_B - \psi_A. \qquad (4)$$

The innate properties of the individual constituent materials (the innate built-in potentials, $\psi_A$ and $\psi_B$) therefore dictates the built-in potential at the interfaces under equilibrium, irrespective of complex interfacial details. In other words, the built-in potential is subject to properties of the interface *only* when the system is not allowed to reach equilibrium. This sheds an interesting light on the previously overlooked constructionist view in interface science.

To illustrate the consistency of these definitions with regard to charge transfer at the interface, we perform a series of ab-initio density functional theory (DFT) calculations. We consider 12 different M/M heterojunction interfaces: Al/Ag, Al/Au, Ag/Au and Ni/Cu heterojunctions with interfaces along the (100), (110) and (111) orientations (Supplementary Materials). First, for each case we directly calculate $\Delta\psi$ by the difference in $V^{(m)}$ (Eq. 3) through DFT calculation of relaxed heterostructures. These values are then compared to the value of $\Delta\psi$ calculated from the newly-defined innate built-in potentials (Eq. 4). As shown in fig. S6A, we see essentially exact agreement. For comparison, we also consider the work function difference ($\varphi_B - \varphi_A$) to estimate $\Delta\psi_{A|B}$ (fig. S6A). We find that the work function difference gives at best a rough prediction for the built-in potential. The failure of the work function-based approach can be understood because it is inappropriate to describe interfacial properties using work functions which describe charge relaxations into vacuum. In addition to comparing Eqs. 3 and 4, we have also investigated the built-in potential through charge density-based methods and found a full consistency of our approach (see Supplementary Materials).

**Metal/semiconductor interface**

When electronic equilibrium of a semiconductor can be established, e.g. by external doping, the built-in potentials at metal/semiconductor (M/S) interfaces are uniquely defined. It should be noted that in a doped semiconductor the *innate built-in potential* (Eq. 2) is, strictly speaking, not an intrinsic property of the semiconductor, but is a *bulk* property which includes the effects of doping. For such cases, our theory above has important implications. Fig. 3A illustrates the band diagrams of the ideal M/S junctions with *n*-doped semiconductors, where $V^{(m)}$ is taken as the energy reference. Much of the previous efforts in understanding the interfacial dipole at M/S interfaces has been focused on the details of the chemistry at the interface (*15,18*). However, we find that interfacial chemistry has *no* effect on the amount of the total electric dipole across the interface (Figs. 3C,D). It is because Eq. 4 guarantees that the built-in potential is the sole source for the dipole as long as the Fermi levels are aligned. The perceived-to-be-local interfacial effects such as chemical bonding, defects, and disorder at the immediate interface are thus all associated with higher-order electrostatic contributions, which contain contributions away from the immediate interface. This is illustrated in Fig. 3D, where the higher order contributions do *not* cause a rigid potential shift on each side of the heterojunctions (black dashed lines), instead the consequence is the development of a potential barrier (or well), $\Delta$ in the semiconductor region in Fig. 3. We should note that the higher order contributions also depend on the semiconductor bulk properties, because the dopant concentration of the bulk influences its surface chemical activity, the ability of the subsequent interfacial states to hold electrons, and the depletion width (*12,15*). Such a decomposition at the interface offers an opportunity to quantify, for example, the dopant concentration dependent Schottky barriers (SBs) which has been neglected under the flat band condition (*12*).

Another important advantage of introducing the innate built-in potential is that it provides a clear distinction between innate and non-innate (i.e., interfacial) properties of heterojunctions, which has long been ambiguous, to enable a truly quantitative discussion on the mystery of SBs and band offsets (*15-21,27-34*). In particular, *n*-type SB, $\Phi_{n\text{-SB}}$, is the barrier electrons need to overcome in order to transport from metal to semiconductor. In the Schottky-Mott model (*35,36*), this becomes $\Phi_{n\text{-SB}}^{\text{S-M}} = \varphi_{\text{M}} - \chi_{\text{S}}$, where $\varphi_{\text{M}}$ is the work function of metal and $\chi_{\text{S}}$ is the electron affinity of semiconductor. Here, we define the innate *n*-type SB which depends solely on bulk properties, $\Phi_{n\text{-SB}}^{0} = \psi_{\text{M}} - \chi_{\text{S}}^{\text{innate}}$, where $\psi_{\text{M}}$ is the innate built-in potential of metal and $\chi_{\text{S}}^{\text{innate}}$ is the innate electron affinity of semiconductor (see Fig. 3A). The difference between $\Phi_{n\text{-SB}}$ and $\Phi_{n\text{-SB}}^{0}$ ($\Delta = \Phi_{n\text{-SB}} - \Phi_{n\text{-SB}}^{0}$ in Fig. 3) can then be completely attributed to interfacial chemistry. When $\Delta \approx 0$, for example in van der Waals (vdW) systems (*37,38*), one can expect that the SBs should be determined from bulk properties. For such systems, the Schottky-Mott model is also expected to yield a good approximation of the SB, as has been recently shown in Ref. *39*. We have calculated *n*-type SBs for 1T-MS$_2$/2H-MoS$_2$ vdW heterostructures with M = (Ti, V, Nb, Mo, Ta, W). As shown in fig. S7, our newly-defined $\Phi_{n\text{-SB}}^{0}$ yields a systematic improvement over the Schottky-Mott model.

**Metal/aqueous interface**

In metal/aqueous interfaces, electronic equilibrium is not reached (*40*) in contrast to M/M or M/S interfaces. The generality of Eq. 3 allows us to construct a unified theory that not only includes solid/solid interfaces but can even be used to describe the open circuit voltage in electrochemical circuits which include metal/aqueous interfaces. Consider a galvanic circuit $\text{M}_{\text{m}}|\text{M}_1|aq|\text{M}_2|\text{M}_{\text{m}}^{'}$, where the two electrodes (M$_1$ and M$_2$) are separated by an aqueous solution (*aq*), and the cell potential is measured by an instrument consisting metal $\text{M}_{\text{m}} = \text{M}_{\text{m}}^{'}$ (Fig. 4A). Although the open circuit voltage, $E_{\text{OCV}}$, measures the sum of all the built-in potentials developed in the cell, $E_{\text{OCV}} = \sum_{n} \Delta \psi_{n} = \Delta \psi_{\text{M}_{\text{m}}|\text{M}_1} + \Delta \psi_{\text{M}_1|aq} + \Delta \psi_{aq|\text{M}_2} + \Delta \psi_{\text{M}_2|\text{M}_{\text{m}}^{'}}$, it has not been possible to determine the built-in potentials at the individual interfaces (*13,14*). Using Eq. 3, however, we can quantify the built-in potential at each interface. In Fig. 4B, we find that our theory predicts the correct $E_{\text{OCV}}$, which is the Fermi level difference between the two electrodes. Moreover, the previously thought-to-be ill-defined Galvani potentials (or inner potentials) for each material in the circuit can also be determined explicitly from the maximum values of the planar averaged electrostatic potentials $V^{(\text{m})}$ (see red and black curves in Fig. 4B). This opens a way to quantitatively design galvanic circuits and, more broadly, other multicomponent electrochemical devices. The $E_{\text{OCV}}$ and the built-in potentials can be calculated with a proper modelling of the liquid phase, for example, by using molecular dynamics method (*40-42*) and space-time averaging method (*43*) within DFT. Note that in homogeneous liquid and gas phases the maximum level of the planar averaged electrostatic potential is always equal to its average potential, $V^{(\text{m})} = V_0$, which leads to $B = 0$ and $D = S$.

The determination of the built-in potential also leads to a rigorous definition of the single electrode potentials that determine $E_{\text{OCV}}$ in a galvanic circuit (*13,40,44*). Using Eq. 4 at metal/metal interfaces, $E_{\text{OCV}}$ can be rewritten in terms of single electrode potentials,

$$E_{\text{OCV}} = \left(\psi_{M_1} + \Delta\psi_{M_1|aq}\right) - \left(\psi_{M_2} + \Delta\psi_{M_2|aq}\right) = E_{M_1|aq} - E_{M_2|aq}, \text{ where } E_{M_1|aq} \text{ and } E_{M_2|aq} \text{ are}$$

respectively the single electrode potential of the two electrodes. We define the *single electrode potential* of an electrode $M$ as

$$E_{M|aq} = \psi_M + \Delta\psi_{M|aq}, \tag{5}$$

where $\psi_M$ and $\Delta\psi_{M|aq}$ are respectively given by Eqs. 2 and 3. Notice that $\Delta\psi_{M|aq}$ (and consequently $E_{M|aq}$) includes non-innate properties, because the M/*aq* interface is not in electronic equilibrium. By our definition, $E_{M|aq}$ is the energy difference between $V_{aq}^{(m)}(=V_0)$ in aqueous solution and the Fermi level of electrode (Fig. 4B). This may be compared to the widely used "absolute electrode potential" suggested by Trassati (*40,44*), $E_{M|aq}^{ab}$, which is defined as the energy difference between vacuum near the solution and the Fermi level of electrode (fig. S8). Although $E_{M|aq}^{ab}$ is convenient to measure (*13,44*), such a definition includes surface properties of the solution. On the other hand, the single electrode potential $E_{M|aq}$, Eq. 5, is explicitly determined independent of any other surfaces and interfaces (see Fig. 4B and fig. S8). To the best of our knowledge, this is the first time a rigorous quantitative definition of the single electrode potential has been given.

**Conclusions**

We have presented a rigorous definition of the built-in potential that provides a unified description of the band alignments at interfaces between all classes of matter. It enables a clear distinction between innate and non-innate properties at interfaces, allowing one to construct quantitative models which unambiguously describe a wide range of surface/interfacial properties and phenomena. Our findings suggest that, under electronic equilibrium, the constructionist hypothesis *is* valid. While this condition is always true when free carriers are present, for large gap semiconductors or electrolytes the constructionist view breaks down because electronic equilibrium cannot be reached. Beyond the interfacial physics, the insight into the common energy reference of bulk solids provides clues for understanding the long-standing problem of defining an average electrostatic potential of infinitely large systems.

**Acknowledgments:** DHC thanks J. Bang, B. Ryu, and Y. J. Oh for their fruitful comments. This work was supported by the US DOE Grant No. DE-SC0002623. The supercomputer time sponsored by NERSC under DOE contract No. DE-AC02-05CH11231 and the CCI at RPI are also acknowledged.

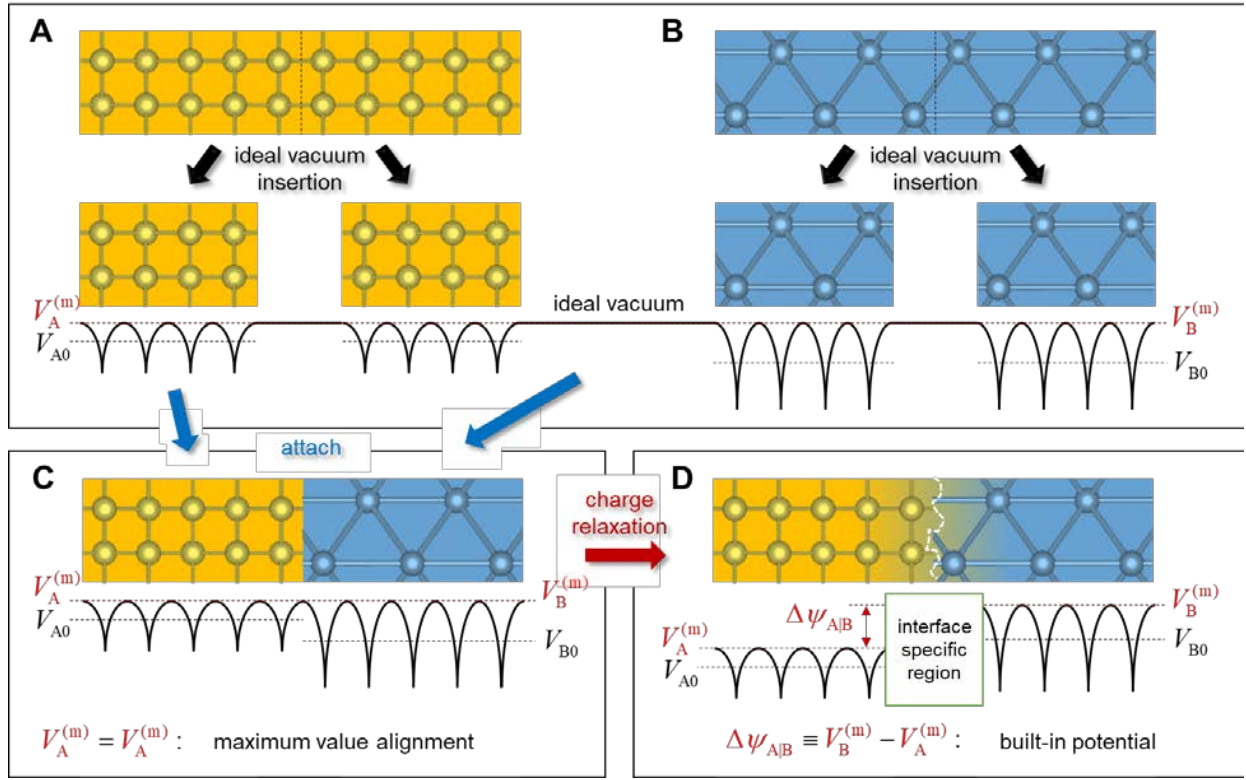

**Fig. 1. Schematic diagram for the interface generation and the built-in potential. (A-B)** Illustration for the vacuum insertion within the bulk materials A and B. The maximum values ($V_A^{(m)}$ and $V_B^{(m)}$) and the average potential ($V_{A0}$ and $V_{B0}$) of the planar averaged electrostatic potentials are denoted as red and black dashed lines, respectively. **(C)** The ideal A/B interface obtained by attaching the two ideal surfaces. The maximum values ($V_A^{(m)}$ and $V_B^{(m)}$) are naturally aligned to the same energy, the *ideal vacuum* level, before the charge relaxation (see the text for more details). **(D)** The A/B interface after charge relaxation. The built-in potential $\Delta \psi_{A|B}$ is well-defined even though the boundary between two dissimilar materials is ambiguous.

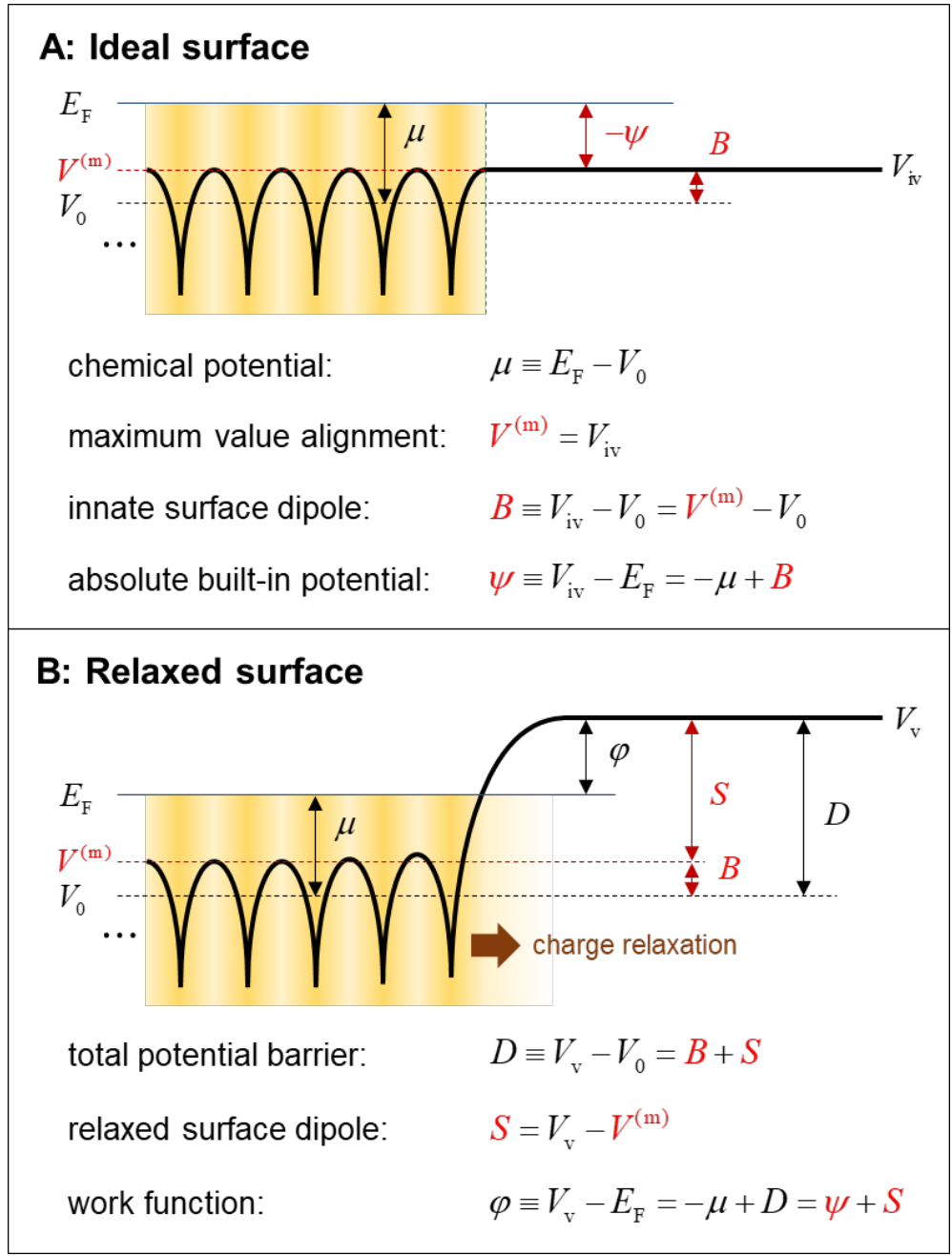

**Fig. 2. Band diagram near solid/vacuum interfaces.** (**A**) An ideal solid surface. The average electrostatic potential $V_0$ and the Fermi level $E_F$ are denoted as black dashed line and dark blue solid line, respectively. The maximum value $V^{(m)}$ is aligned to the *ideal vacuum* $V_{iv}$ (proof given in the Supplementary Materials). (**B**) Solid surface after charge relaxation. The total potential barrier $D$ is the difference between $V_0$ and the vacuum $V_v$.

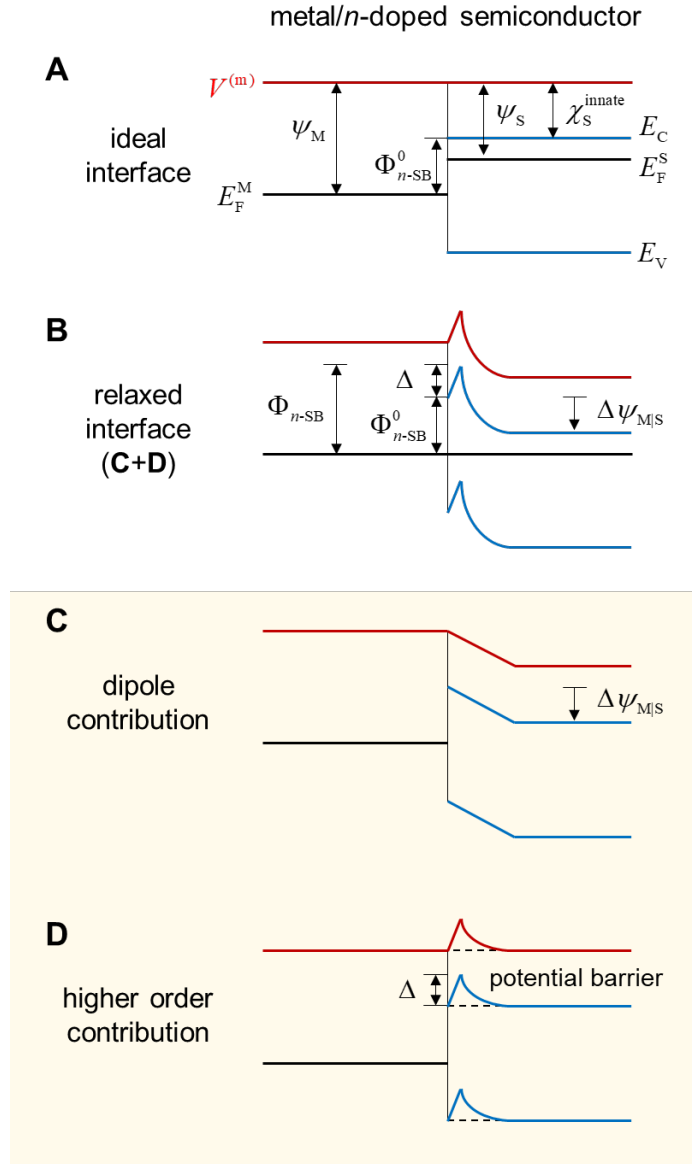

**Fig. 3. Schematic band diagram for metal/semiconductor heterojunctions.** (**A**) Depicts the innate alignment of the ideal M/S interface for the case of an *n*-doped semiconductor without any electronic or ionic relaxation. The common energy reference, $V^{(m)}$, is denoted by the red solid lines. The Fermi levels and the band edges are denoted as solid black and blue lines, respectively. As there is no relaxation, the Fermi levels are not aligned. (**B**) The M/S interfaces after allowing for full ionic and electronic relaxations. The full interfacial relaxation depicted in (**B**) can be decomposed into dipole and higher order contributions, which are depicted in (**C**) and (**D**), respectively. Accordingly, the *n*-type SB $\Phi_{n\text{-SB}}$ is decomposed into $\Phi^0_{n\text{-SB}}$ and $\Delta$. The dipole relaxation, shown in (**C**), leads to a rigid shift in the band edges determining the built in potential ($\Delta\psi_{M/S} = \psi_M - \psi_S$). Higher order contributions are shown in (**D**) and are associated with a potential barrier developed in the semiconductor region ($\Delta$) which does not alter the long range band alignment.

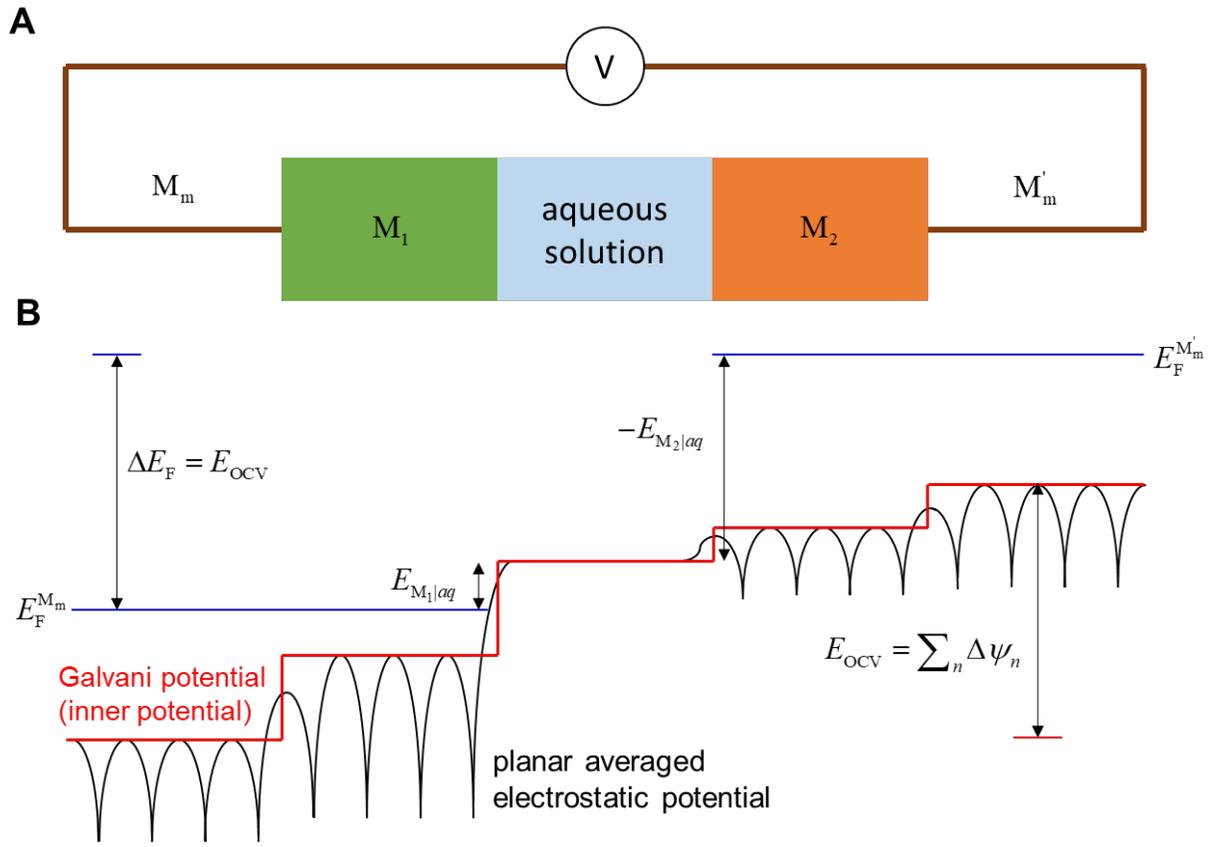

**Fig. 4. A galvanic circuit** $M_m|M_1|aq|M_2|M_m'$. **(A)** Schematic device setup and **(B)** the potential profile across the circuit. Each maximum value of the planar averaged electrostatic potentials in bulk regions of the individual materials (black curve) determines the Galvani potential of the each material (red curve). The built-in potentials at the individual interfaces, $\Delta\psi_n$, are the differences in the Galvani potentials. The open circuit voltage, $E_{OCV}$, is equal to the Fermi level difference, the sum of all the total built-in potentials, and the single electrode potential difference ($E_{OCV} = \Delta E_F = \sum_n \Delta\psi_n = E_{M_1|aq} - E_{M_2|aq}$).